\documentclass[amsmath,amssymb,prl,final,showpacs,twocolumn]{revtex4}
\usepackage{bm,hyphenat,xspace}
\usepackage{graphicx,epsfig}

\newcommand{\scl}{0.63}
  
\newcommand{\Eq}{Eq.}

\newcommand{\Fig}{Fig.}

\newcommand{\Ref}{Ref.}
\newcommand{\Refs}{Refs.}

\newcommand {\mbf}[1]{{\mathbf{#1}}}

\newcommand {\mcu}{\mathcal{U}}

\newcommand{\cm}{\mathrm{c\!\:\!.m\!\:\!.}}

\newcommand{\He}{{}^3\mathrm{He}}

\newcommand{\nH}{n\text{-}{}^3\mathrm{H}}
\newcommand{\pHe}{p\text{-}{}^3\mathrm{He}}

\newcommand{\pd}{p\text{-}d}
\newcommand{\nd}{n\text{-}d}

\begin{document}
 
\title {Four-body calculation of proton-$\He$ scattering}
 
\author{A.~Deltuva} 
%\email{deltuva@cii.fc.ul.pt}
\affiliation{Centro de F\'{\i}sica Nuclear da Universidade de Lisboa, 
P-1649-003 Lisboa, Portugal }

\author{A.~C.~Fonseca} 
\affiliation{Centro de F\'{\i}sica Nuclear da Universidade de Lisboa, 
P-1649-003 Lisboa, Portugal }

\received{November 10, 2006}

\begin{abstract}
The four-body equations of Alt, Grassberger and Sandhas are solved,
for the first time, for proton-${}^3\mathrm{He}$ scattering including 
the Coulomb interaction  between the three
protons using the method of screening and renormalization as it was done
recently for proton-deuteron scattering. 
Various realistic two-nucleon potentials 
%including the one derived from chiral perturbation theory 
are used. Large Coulomb effects are seen on all observables.
Comparison with data at different energies shows
large deviations in the proton analyzing power but quite
reasonable agreement in other observables.    
The effect of nucleon-nucleon magnetic moment interaction
and correlations between $\pd$ and $\pHe$ analyzing powers are studied.

\end{abstract}

 \maketitle

%%%%%%%%%%%%%%%%%%%%%%%%%%%%%%%%%%%%%%%%%%%%%%%%%%%%%%%%%%%%%%%%%%%%%%%%%%%%%%%

Modern calculations of light nuclear systems $A \leq 12$ are essential to
our understanding of the force models that have been developed to describe how
nucleons interact at low energies~\cite{pieper:01a,pieper:02a}. 
Of these nuclear systems, the four-nucleon $(4N)$ system is particularly 
important because it gives
rise, experimentally, to the simplest set of nuclear reactions that shows the
complexity of heavier systems and the Coulomb interaction manifest  itself in
new ways relative to what is observed in the three-nucleon $(3N)$ system.
Theoretically it is also important because with powerful 
numerical techniques and fast computers one can calculate not only bound state
properties~\cite{nogga:02a} but also scattering observables
\cite{viviani:98a,cieselski:99,fonseca:99a,viviani:01a,fonseca:02a,%
lazauskas:04a,lazauskas:05a} for a
number of elastic, transfer and breakup reactions that place new challenges to
our understanding of the underlying force models. The importance of scattering
calculations also has to do with the possibility to probe states in the
continuum associated with specific resonances, states of higher angular
momentum than corresponding bound states, effects that depend on the spin
orientation of the projectile and/or target, and threshold effects on the
observables, among others. 

While the three-nucleon system has been extensively
studied~\cite{gloeckle:96a,golak:05a} through neutron-deuteron $(nd)$ and 
proton-deuteron $(pd)$ elastic scattering and breakup experiments, 
exact calculations
using realistic force models as well as interactions derived from Effective
Field Theory were restricted, for a long time, to the $nd$ system due to
limitations in including the Coulomb force in the description of $pd$ 
scattering beyond low energy  $pd \to pd$ and $pd \leftrightarrow
\gamma\He$ calculations \cite{kievsky:01a,viviani:00a} in the
framework of the variational hyperspherical approach.
The situation has now changed due to the work of 
Refs.~\cite{deltuva:05a,deltuva:05c}
where calculations of $pd \to pd$, $pd \to ppn$, 
$pd \leftrightarrow \gamma \He$, $\gamma \He \to ppn$, 
$e \He \to e' pd$, and $e \He \to e' ppn$
were performed at energies ranging from 1 MeV in the center of mass (c.m.)
system to the pion production threshold. The work is based on the
solution of the momentum-space Alt, Grassberger and Sandhas (AGS)
equations~\cite{alt:67a} together with the screening and renormalization
approach \cite{taylor:74a,alt:78a,alt:80a} for the Coulomb interaction 
leading to the results of observables that are independent of the screening
radius, provided it is sufficiently large.

In the present manuscript for the first time we extend the method 
of Refs.~\cite{deltuva:05a,deltuva:05c} to the 
$\pHe$ elastic scattering using the four-body AGS
equations~\cite{grassberger:67}. The aim is to bring the $4N$
scattering problem to the same level of understanding in terms of the
underlying two-nucleon $(2N)$
forces as already exists for $3N$, which means that
calculations are carried out without approximations on the $2N$ 
transition matrix (t-matrix) like in Ref.~\cite{fonseca:99a} 
or limitations on the choice of basis functions as in
Refs.~\cite{pfitzinger:01a,fisher:06}. 
Therefore, after partial wave decomposition, the AGS
equations are three-variable integral equations that are solved numerically
without any approximations beyond the usual discretization of continuum
variables on a finite momentum mesh. The results we present here are
converged vis-a-vis number of partial waves and momentum meshpoints
as well as the value of the screening radius of the Coulomb potential.
These calculations are also an extension to
$\pHe$ of the work already developed for $\nH$~\cite{deltuva:07a}, and were
presented for the first time in Ref.~\cite{fonseca:fb18}. Our work follows 
the work of Refs.~\cite{viviani:01a,fisher:06,pfitzinger:01a}, 
but with greater number of $2N$, $3N$, and $4N$ partial waves 
in order to get fully  converged results for the spin observables
and with various $2N$ potentials. The advantage of the present work is that
it is easier to extended to inelastic reactions 
and to use with nonlocal interactions.

Our description of $4N$ scattering is based on the symmetrized 
four-body AGS equations given in Ref.~\cite{deltuva:07a} where 
the solution technique is discussed in detail.
In order to include the Coulomb interaction we follow the methodology 
of \Refs~\cite{deltuva:05a,deltuva:05c}
and add to the nuclear $pp$ potential
the screened Coulomb one $w_R$ that, in configuration space, is given by
\begin{equation} \label{eq:wr}
w_R(r) = w(r) \, e^{-(r/R)^n},
\end{equation}
where  $w(r) = \alpha_e/r$ is the true Coulomb potential, 
$\alpha_e \simeq 1/137$ is the fine structure constant, 
and $n$ controls the smoothness of the screening; $n=4$ is the optimal 
value which ensures that $w_R(r)$ approximates well $w(r)$ for $r < R$
and simultaneously vanishes rapidly for $r>R$,
providing a comparatively fast convergence of the partial-wave expansion.
The screening radius $R$ must be considerably 
larger than the range of the strong interaction but from
the point of view of scattering theory $w_R$ is still of short range.
Therefore the equations  of Ref.~\cite{deltuva:07a} become $R$ dependent.
The transition operators $\mcu^{\alpha\beta}_{(R)}$ where
$\alpha(\beta) = 1$ and 2 corresponds to initial/final  
$1+3$  and $2+2$ two-cluster states, respectively, satisfy the symmetrized
AGS equations
\begin{subequations}
\begin{align}  \nonumber
\mcu^{11}_{(R)}  = {}& -(G_0 \, t^{(R)}  G_0)^{-1}  P_{34} 
 - P_{34} \, U^1_{(R)}\, G_0 \, t^{(R)} G_0 \, \mcu^{11}_{(R)} \\
& + {U}^2_{(R)}   G_0 \, t^{(R)} G_0 \, \mcu^{21}_{(R)} , 
\label{eq:U11} \\  \nonumber
\mcu^{21}_{(R)}  = {}&  (G_0 \, t^{(R)}  G_0)^{-1} \, (1 - P_{34})
\\ & + (1 - P_{34}) U^1_{(R)}\, G_0 \, t^{(R)}  G_0 \,
\mcu^{11}_{(R)} . \label{eq:tildeU21}
\end{align}
\end{subequations}
Here $G_0$ is the four free particle Green's function and $t^{(R)}$ the 
two-nucleon t-matrix derived from nuclear potential plus screened 
Coulomb between $pp$ pairs. The  operators $U^\alpha_{(R)}$ obtained from    
\begin{subequations} 
\begin{align}
\label{eq:U}
U^{\alpha}_{(R)} = {} & P_\alpha G_0^{-1} + 
P_\alpha \, t^{(R)}\, G_0 \, U^{\alpha}_{(R)}, \\
\label{eq:P}
P_1 = {} & P_{12}\, P_{23} + P_{13}\, P_{23}, \\
\label{eq:tildeP} 
P_2 = {} & P_{13}\, P_{24}, 
\end{align}
\end{subequations}
are the symmetrized AGS operators for the $1+(3)$ and $(2)+(2)$ subsystems
and $P_{ij}$ is the permutation operator of particles $i$ and $j$.
Defining the initial/final $1+(3)$ and $(2)+(2)$ states with relative
two-body momentum $\mbf{p}$
\begin{gather} \label{eq:phi1}
  | \phi_\alpha^{(R)} (\mbf{p}) \rangle = 
G_0 \, t^{(R)}  P_\alpha | \phi_\alpha^{(R)} (\mbf{p}) \rangle ,
\end{gather}
the amplitudes for $1+3\to1+3$ and $1+3\to2+2$ are obtained as
$\langle \mbf{p}_f | T^{\alpha\beta}_{(R)} | \mbf{p}_i \rangle = 
S_{\alpha\beta} \langle \phi_\alpha^{(R)} (\mbf{p}_f) | 
\mcu^{\alpha\beta}_{(R)} |\phi_\beta^{(R)} (\mbf{p}_i)\rangle $ with
$S_{11} = 3$ and $S_{21} = \sqrt{3}$.

In close analogy with $pd$ elastic scattering, the full scattering amplitude,
when calculated between initial and final $\pHe$ states,
 may  be decomposed as follows
\begin{equation}\label{eq:U11R}
T^{11}_{(R)} = t_R^{\cm} + [T^{11}_{(R)} - t_R^{\cm}],
\end{equation}
with the long-range part $t_R^{\cm}$ being the two-body t-matrix derived 
from the screened Coulomb potential of the form \eqref{eq:wr} between the 
proton and the c.m.  of $\He$, and the remaining Coulomb-distorted 
short-range part $[T^{11}_{(R)} - t_R^{\cm}]$ as demonstrated in 
\Refs~\cite{alt:80a,deltuva:prep}.
Applying the renormalization procedure, i.e., multiplying both sides 
of \Eq~\eqref{eq:U11R} by
the renormalization factor $Z_R^{-1}$ \cite{deltuva:05a,alt:80a},
in the $R \to \infty$ limit, yields the full $1+3 \to 1+3$ transition 
amplitude in the presence of Coulomb
\begin{gather}      \label{eq:T} 
  \begin{split}
    \langle \mbf{p}_f| T^{11} |\mbf{p}_i \rangle  = {} &
    \langle \mbf{p}_f| t_C^{\cm} |\mbf{p}_i \rangle   \\
    &+ \lim_{R \to \infty} \left\{ 
    \langle \mbf{p}_f| [ T^{11}_{(R)} - t_R^{\cm} ]|\mbf{p}_i \rangle
    Z_R^{-1}  \right\},
    \end{split}
\end{gather}  
where  the  $Z_R^{-1} \langle \mbf{p}_f| t_R^{\cm} |\mbf{p}_i \rangle$
converges (in general, as a distribution) to the 
exact Coulomb amplitude $\langle \mbf{p}_f| t_C^{\cm} |\mbf{p}_i \rangle$ 
between the proton and the c.m. of the $\He$ nucleus,
and therefore is replaced by it. 
The renormalization factor is employed in the
partial-wave dependent form as in \Ref~\cite{deltuva:05a}
\begin{gather} 
\label{eq:ZR}
Z_R =  e^{- 2 i ( \sigma_L - \eta_{LR}) }
\end{gather}
with the diverging screened Coulomb $\pHe$ phase shift $\eta_{LR}$
corresponding to standard boundary conditions
and the proper Coulomb one $\sigma_L$ referring to the
logarithmically distorted proper Coulomb boundary conditions.
The second term in \Eq~\eqref{eq:T}, 
after renormalization by $Z_R^{-1}$, represents the Coulomb-modified
nuclear short-range amplitude. It has to be calculated numerically,
but, due to its short-range nature, the $R \to \infty$ limit
is reached with sufficient accuracy at finite screening radii $R$.
As in $pd$ elastic scattering \cite{deltuva:05a}
one needs larger values of $R$ for decreasing
proton energies, making the convergence of the results more
difficult to reach. Nevertheless for $E_p > 2$ MeV the method leads to
very precise results as we demonstrate in \Fig~\ref{fig:R} for the
differential cross section $d\sigma/d\Omega$, proton analyzing power $A_y$,
and $\pHe$ spin correlation coefficient $C_{yy}$ at proton lab energy 
$E_p = 4$ MeV. 
The observables are shown as functions of the c.m. scattering angle.
Fully converged results are obtained with
$R = 12$ fm, but already $R = 8$ and 10 fm results are very close to them.
The calculations include isospin-singlet $2N$ partial waves with 
total angular momentum $I \leq 4$ and  isospin-triplet $2N$ partial waves 
with orbital angular momentum $l_x \leq 7$, $3N$ partial waves
with spectator orbital angular momentum $l_y \leq 7$ and 
total angular momentum $J \leq \frac{13}{2}$, $4N$ partial waves
with 1+3 and 2+2 orbital angular momentum $l_z \leq 7$
and all initial/final $\pHe$ states with orbital angular momentum $L \leq 3$.
The charge-dependent (CD) Bonn potential \cite{machleidt:01a} is used. 
The effect of Coulomb is large in the whole angular
region, particularly for $A_y$ where it reduces the magnitude
of the maximum. The $R = 0$ fm curve corresponds to the so-called
Doleschall approximation which clearly fails to reproduce the
full Coulomb effect.

\renewcommand{\scl}{0.62}
\begin{figure}[!]
\begin{center}
\includegraphics[scale=\scl]{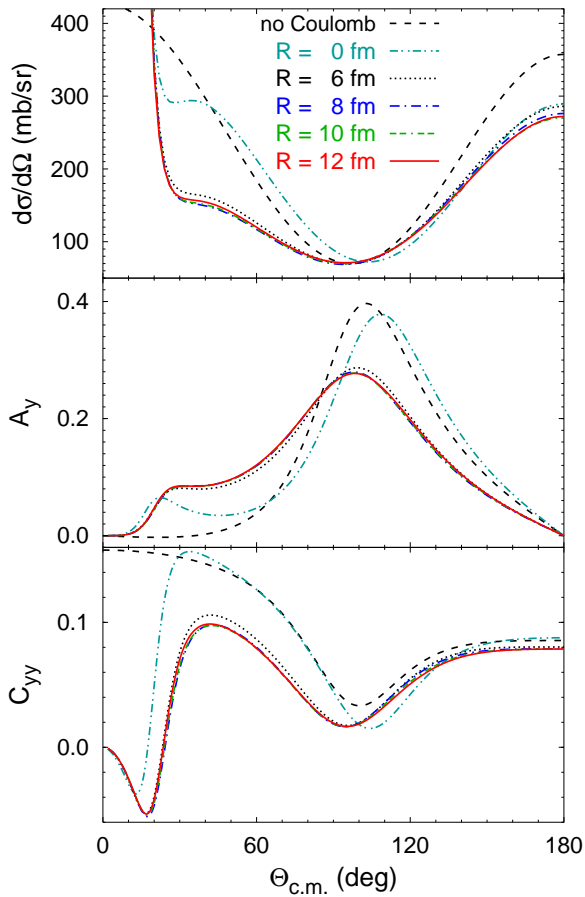}
\end{center}
\caption{\label{fig:R} (Color online)
Convergence of the $\pHe$ scattering  observables with screening radius $R$.
Results for the differential cross section, proton analyzing power $A_y$, and
$\pHe$ spin correlation coefficient $C_{yy}$ at 4~MeV proton lab energy
obtained with screening radius $R=0$~fm (dashed-double-dotted curves),
6~fm (dotted curves), 8~fm (dashed-dotted curves), 
10~fm (double-dashed-dotted curves), and 12~fm (solid curves) are compared.
Results without Coulomb (dashed curves)
are given as reference for the size of the Coulomb effect.}
\end{figure}

\begin{figure}[!]
\begin{center}
\includegraphics[scale=0.58]{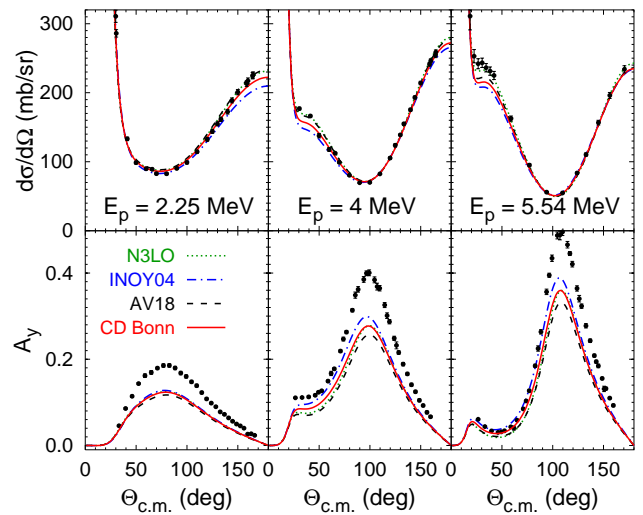}
\end{center}
\caption{\label{fig:dcsA} (Color online)
The differential cross section and proton analyzing power $A_y$
at 2.25, 4.0, and 5.54~MeV proton lab energy.
Results including the Coulomb interaction obtained with potentials
CD Bonn (solid curves), AV18 (dashed curves),
INOY04 (dashed-dotted curves), and N3LO (dotted curves) are compared.
The data  are from \Refs~\cite{fisher:06,mcdonald:64,alley:93}.}
\end{figure}

\begin{figure}[!]
\begin{center}
\includegraphics[scale=0.55]{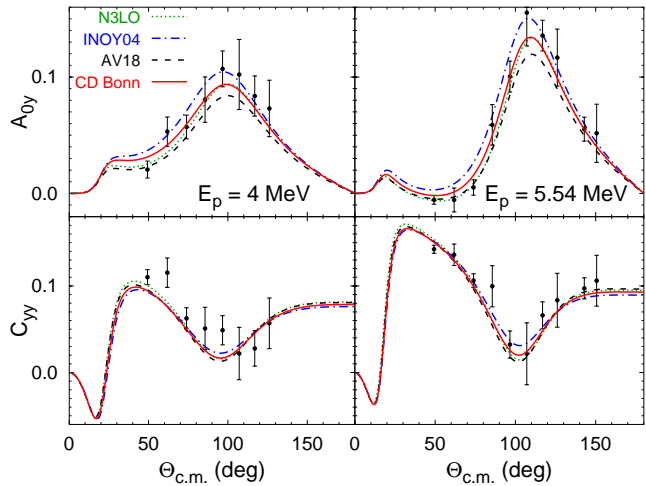}
\end{center}
\caption{\label{fig:A0y} (Color online)
$\He$ target analyzing power $A_{0y}$ and 
spin correlation coefficient $C_{yy}$ at 4.0 and 5.54~MeV proton lab energy. 
Curves as in \Fig~\ref{fig:dcsA}.
The data  are from \Ref~\cite{alley:93}.}
\end{figure}

\begin{figure}[!]
\begin{center}
\includegraphics[scale=0.55]{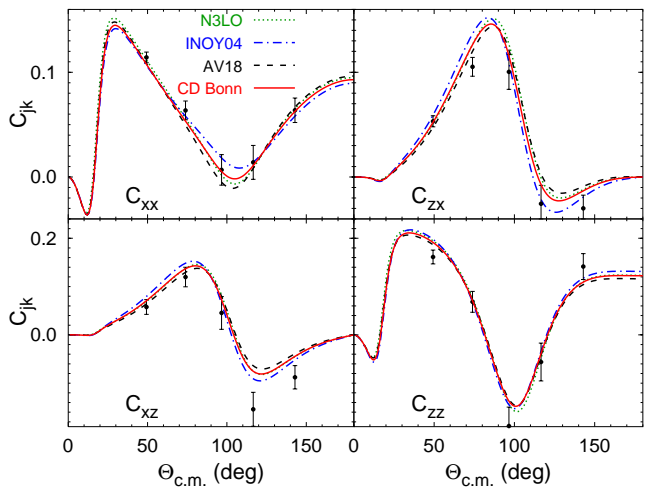}
\end{center}
\caption{\label{fig:Cij} (Color online)
$\pHe$ spin correlation coefficients at 5.54~MeV proton lab energy.
Curves as in \Fig~\ref{fig:dcsA}.
The experimental  data  are from \Ref~\cite{alley:93}.}
\end{figure}

In Figs.~\ref{fig:dcsA} --- \ref{fig:Cij} we compare the results of our 
calculations with data for a number of observables at $E_p = 2.25$,  
4.0, and 5.54 MeV. In addition to CD Bonn
we use AV18 \cite{wiringa:95a}, inside-nonlocal outside-Yukawa (INOY04) 
potential by Doleschall \cite{doleschall:04a,lazauskas:04a}, and 
the one  derived from chiral perturbation theory at 
next-to-next-to-next-to-leading order (N3LO) \cite{entem:03a}.
The $\He$ binding energy (BE) calculated with AV18, N3LO, CD Bonn, and INOY04
potentials is 6.92, 7.13, 7.26, and 7.73 MeV, respectively;
the experimental value is 7.72 MeV. As  in $\nH$ scattering 
\cite{deltuva:07a}, $\pHe$ observables depend on the choice of potential;
predictions with N3LO and AV18 agree best with the cross section data
but it is INOY04 that provides the highest $A_y$ at the peak.
If one considers AV18, CD Bonn, and INOY04 potentials alone, one might be
tempted to conclude about a possible correlation between observables and 
$\He$ BE. Nevertheless, as discussed in Ref.~\cite{deltuva:07a}, N3LO,
for reasons not yet fully understood, breaks this correlation
in the considered energy region. As found in Ref.~\cite{deltuva:07a},
$4N$ $S$-wave phase shifts correlate with the $3N$ BE
\cite{deltuva:07a} but as the energy increases $4N$ $P$-waves become
very important as well and behave differently depending on the choice 
of potential.  Therefore correlations between $\pHe$ 
observables and $\He$ BE cannot be established
easily without further studies, e.g., inclusion of a $3N$ force.

As shown in Figs.~\ref{fig:A0y} --- \ref{fig:Cij} 
$\He$ target analyzing power $A_{0y}$ and $\pHe$ spin correlation 
coefficients $C_{jk}$ are described quite satisfactorily.
This updates the findings of Ref.~\cite{fisher:06} based on AV18 potential
where significant discrepancies were observed for $A_{0y}$ and $C_{yy}$.
However, the proton analyzing power is clearly underestimated
by all potentials.  In contrast to low-energy
$pd$ elastic scattering where variations of the $2N$ interaction
at the maximum of $A_y$ lead to 10\% fluctuations, here we get 15\%, 
which means that the $4N$ system is more sensitive to off-shell differences 
of the $2N$ force than the $3N$ system. 

\begin{figure}[!]
\begin{center}
\includegraphics[scale=0.54]{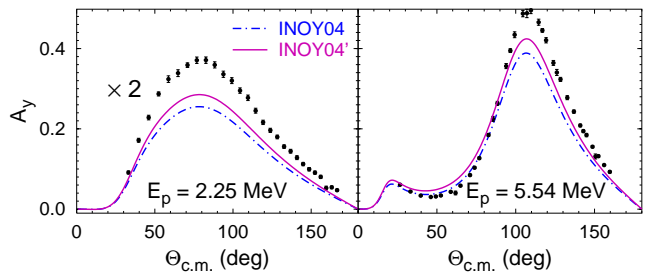}
\end{center}
\caption{\label{fig:Pmod} (Color online)
Proton analyzing power $A_{y}$ at 2.25 and 5.54~MeV proton lab energy. 
Results for the potentials INOY04 (dashed-dotted curves)
and INOY04' (solid curves) are compared.
The experimental  data  are from \Refs~\cite{fisher:06,alley:93}.}
\end{figure}

In Fig.~\ref{fig:Pmod} we compare $A_y$ for potential INOY04 and its version
INOY04' \cite{doleschall:04a,lazauskas:04a} 
with modified $2N$ ${}^3P_I$ wave parameters such that it provides
quite satisfactory description of $A_y$ in low-energy
$\nd$ and $\pd$ scattering at the cost of being inconsistent with the 
$2N$ data. However, for $\pHe$ $A_y$ disagreement with data still persists.

Finally, using AV18 potential
we investigate the effect of $2N$ magnetic moment (MM) interaction.
As for $\pd$ scattering \cite{kievsky:04a} it is most visible for
$A_y$ at low energy where at $E_p = 2.25$ MeV it gives rise 
to a 5.3\%  increase towards the data. % as shown in Fig.~\ref{fig:mm}.
At 4 MeV the MM interaction effect is reduced to 2.7\%.

In conclusion, we have been able to obtain \emph{ab initio} four-nucleon
results for $\pHe$ scattering that include the Coulomb interaction
between the protons for different realistic local and nonlocal
$2N$ interactions. The reliability of the screening and renormalization
approach is demonstrated.
The calculations describe existing data quite well except
proton $A_y$ where there is 25 - 40\% discrepancy at the peak.
We find that $4N$ observables are more sensitive than $3N$ observables
to off-shell changes in the $2N$ interaction, and that curing $A_y$ in
low energy $3N$ scattering through changes in the $2N$ ${}^3P_I$ partial
waves still gives rise to a $\pHe$ $A_y$ deficiency.
A visible effect of $2N$ magnetic moment interaction is found
for $A_y$ at very low energy.

A.D. is supported by 
Funda\c{c}\~{a}o para a Ci\^{e}ncia e a Tecnologia (FCT) 
grant SFRH/BPD/14801/2003
and A.C.F. in part by FCT grant POCTI/ISFL/2/275.

%%%%%%%%%%%%%%%%%%%%%%%%%%%%%%%%%%%%%%%%%%%%%%%%%%%%%%%%%%%%%%%%%%%%%%%%%%%%%
%\bibliographystyle{prsty3}
%\bibliography{abbrev,pre80,80-89,90-99,200x,clmb,ad,4N}

\end{document}